\documentclass[aps,prb,showpacs,onecolumn]{revtex4}

\usepackage{amssymb}
\usepackage{amsmath}
\usepackage{bm}
\usepackage{graphics}
\usepackage{natbib}

\begin{document}
\title{Tunable pure spin currents in a triple-quantum-dot ring}
\author{Weijiang Gong}
\author{Yisong Zheng}\email[Correspondence author. Email: ]{zys@mail.jlu.edu.cn}
\author{Tianquan L\"{u}}
\affiliation{Department of physics, Jilin University, Changchun
130023, People's Republic of China}
\date{\today}

\begin{abstract}
Electron transport properties in a triple-quantum-dot ring with
three terminals are theoretically studied. By introducing local
Rashba spin-orbit interaction on an individual quantum dot, we
calculate the charge and spin currents in one lead. We find that a
pure spin current appears in the absence of a magnetic field. The
polarization direction of the spin current can be inverted by
altering the bias voltage. In addition, by tuning the magnetic field
strength, the charge and spin currents reach their respective peaks
alternately.
\end{abstract}
% \keywords{}
\pacs{73.63.Kv, 71.70.Ej, 72.25.-b, 85.75.-d} \maketitle

\bigskip
%\section{Introduction}
One of the central issue in spintronics is how to realize the spin
accumulation and spin transport in nano-devices. Recently, there has
been many theoretical proposals to achieve the pure spin current
without an accompanying charge current in mesoscopic systems, such
as the use of spin Hall effects,\cite{Zhang,Kato} optical spin
orientation by linearly polarized light,\cite{Hbner,Bhat} adiabatic
or nonadiabatic spin pumping in metals and
semiconductors,\cite{WangBG,Watson} generation in three-terminal
spin devices.\cite{Guo} Among these schemes, spin-orbit(SO) coupling
is exploited to influences the electron spin state. In particular,
in low-dimensional structures Rashba SO interaction comes into play
by introducing an electric potential to destroy the symmetry of
space inversion in an arbitrary spatial
direction.\cite{Rashba,Rashba2} Thus, by virtue of the Rashba
interaction, electric control and manipulation of the electronic
spin state is feasible.\cite{Kim,Shelysh,Sun,Loss,Souma}
\par
In this Letter, we introduce Rashba interaction to act locally on
one component quantum dot(QD) of a triple-QD ring with three
terminals. Our theoretical investigation indicates that it is
possible to form the pure spin current in one of the three leads
even in the absence of a magnetic field. And the polarization
direction of the spin current can be inverted by altering the bias
voltage.
\par
%\section{Model and formulae}

\begin{figure}
\centering \scalebox{0.42}{\includegraphics{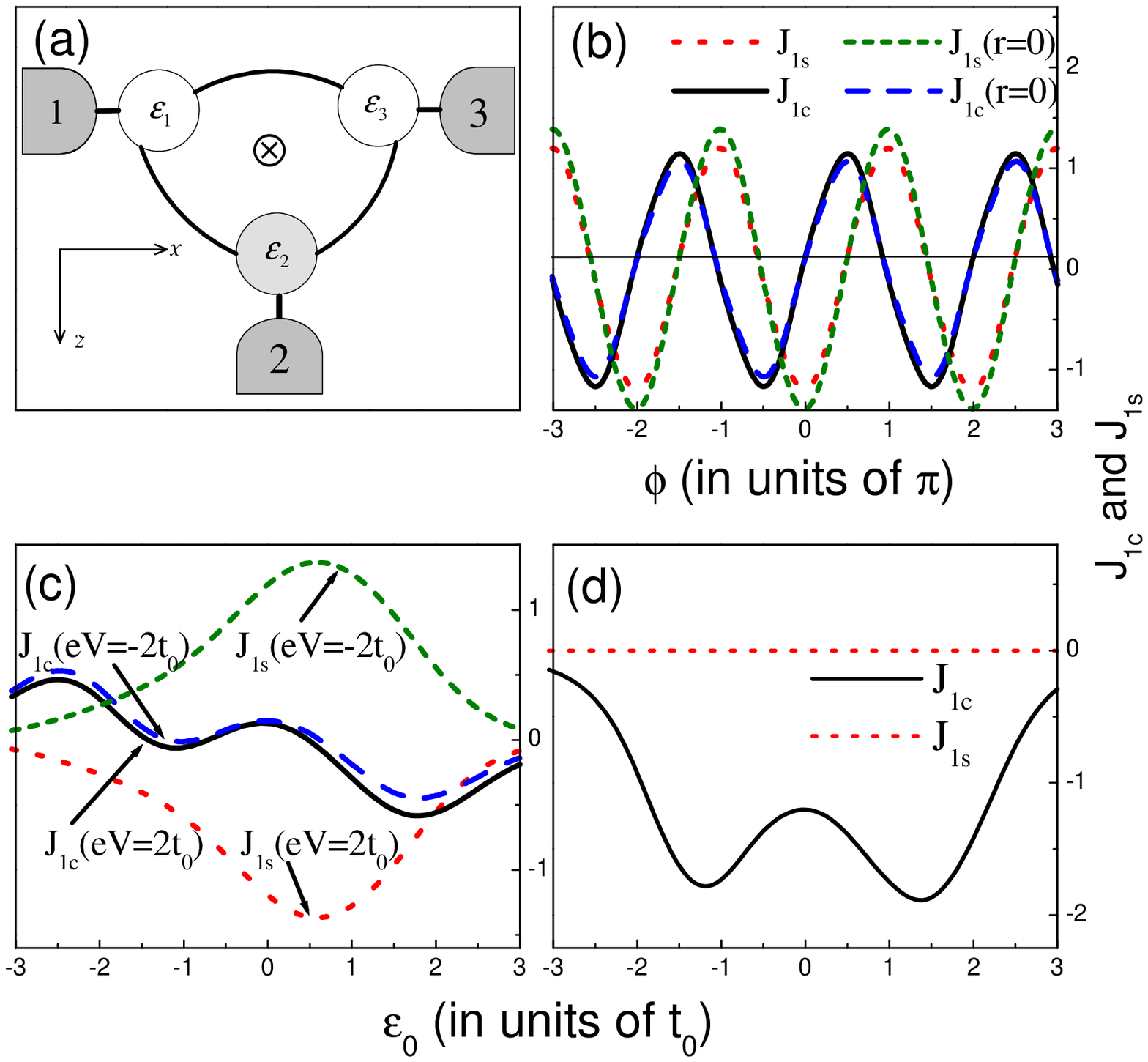}}
\caption{(Color online) (a) A schematic of a three-terminal
triple-QD ring structure with the local Rashba SO interaction on
QD-2. (b)-(c) The currents versus magnetic phase factor $\phi$(b) as
well as the QD level $\varepsilon_0$(c), respectively.
$\Gamma_j=2t_0$, $\tilde{\alpha}=0.5$. In (b) $\varepsilon_j=0$ and
in (c) $\varepsilon_j=\varepsilon_0$, $\phi=0$. The currents without
spin flip terms are shown in (b) for comparison. (d) The currents
versus $\varepsilon_0$ in the two-terminal case($\Gamma_3=0$).
\label{structure}}
\end{figure}

The structure that we consider is illustrated in
Fig.\ref{structure}. The single-particle Hamiltonian for an electron
in such a structure can be written as
$H_s=H_0+H_{so}=\frac{\textbf{P}^2}{2m^*}+V(\textbf{r})+H_{so}$
where, accompanying the kinetic energy term
$\frac{\textbf{P}^2}{2m^*}$, the electron confined potential
$V(\textbf{r})$ defines the structure geometry; And
$H_{so}=\frac{\hat{y}}{2\hbar}\cdot[\alpha(\hat{\sigma}\times
\textbf{p})+(\hat{\sigma}\times \textbf{p})\alpha]$ denotes the
local Rashba SO coupling on QD-2 (QD-j represents the QD with a
single-particle level $\varepsilon_j$ shown in
Fig.\ref{structure}(a)). We select the basis set
$\{\psi_{k_j}\chi_{\sigma},\psi_j\chi_{\sigma}\}$($j$=1,2,3) to
second-quantize the Hamiltonian. The wavefunctions $\psi_j$ and
$\psi_{k_j}$ have the physical meaning of the orbital eigenstates of
the isolated QD and leads, in the absence of Rashba interaction,
where $k_j$ indicates the continuum state in lead-j. $\chi_{\sigma}$
with $\sigma=\uparrow,\downarrow$ denotes the eigenstates of Pauli
spin operator $\hat{\sigma}_z$.
\par
The second-quantized Hamiltonian consists of three parts: ${\cal
H}_{s}={\cal H}_{c}+{\cal H}_{d}+{\cal H}_{t}$.
\begin{eqnarray}
{\cal H}_{c}&=&\underset{\sigma,k_j}{\sum }\varepsilon
_{k_j}c_{k_j\sigma}^\dag c_{k_j\sigma },\notag\\
{\cal H}_d&=&\sum_{j=1, \sigma}^{3}\varepsilon
_{j}d^\dag_{j\sigma}d_{j\sigma}
+\sum_{l=1,\sigma}^{2}[t_{l\sigma}d^\dag_{l\sigma}d_{l+1\sigma}+r_l(d_{l\downarrow}^\dag
d_{l+1\uparrow}\notag\\
&&-d_{l+1\downarrow}^\dag d_{l\uparrow})]+t_{3}e^{i\phi}d^\dag_{3\sigma}d_{1\sigma}+\mathrm {H.c.}\notag\\
{\cal H}_{t}&=&\underset{\sigma k_j }{\sum }V_{j\sigma}
d^\dag_{j\sigma}c_{k_j\sigma}+\mathrm {H.c.},\label{2}
\end{eqnarray}
where $c_{k_j\sigma}^\dag$ and $d^{\dag}_{j\sigma}$ $(
c_{k_j\sigma}$ and $d_{j\sigma})$ are the creation (annihilation)
operators corresponding to the basis states in lead-j and QD-j.
$\varepsilon _{k_j}$ is the single-particle level in lead-j.
$V_{j\sigma}=\langle \psi_j\chi_\sigma|H
_s|\psi_{k_j}\chi_\sigma\rangle$ denotes QD-lead hopping amplitude.
The interdot hopping amplitude, written as $t_{l\sigma}=t_l-i\sigma
s_l$($l=1,2$), has two contributions:
$t_{l}=\langle\psi_{l}|H_0|\psi_{l+1}\rangle$ is the ordinary
transfer integral irrelevant to the Rashba interaction;
$s_l=i\langle\psi_{l}|\alpha p_x+p_x\alpha|\psi_{l+1}\rangle$
indicates the strength of spin precession. In addition, the interdot
spin flip term has the strength $r_l=\langle\psi_{l}|\alpha
p_z+p_z\alpha|\psi_{l+1}\rangle$. A magnetic field penetrating the
ring is described by a geometric phase factor $\phi$.
\par
Without loss of generality, we assume that each QD confines the
electron by an isotropic harmonic potential ${1\over
2}m^*\omega_0r^2$; and the three QDs are positioned on a circle
equidistantly. Then by a straightforward derivation, we find some
rough relationships between the relevant parameters in the above
Hamiltonian: $t_1=t_2$, $s_1=s_2$, $r_1=-r_2$, and
$|s_l|=|r_l|\simeq\tilde{\alpha} t_l$, where
$\tilde{\alpha}=\alpha\sqrt{m^*}/(3\sqrt{\hbar^3\omega_0})$, is a
dimensionless Rashba coefficient. Following these relations we can
rewrite the interdot hopping amplitude in an alternative form:
$t_{l\sigma}=t_l\sqrt{1+\tilde{\alpha}^2}e^{-i\sigma\varphi}=t_0e^{-i\sigma\varphi}$
with $\varphi=\tan^{-1}\tilde{\alpha}$. Thus, just three independent
parameters, $t_0$, $\tilde{\alpha}$ and the magnetic phase factor
$\phi$, are needed to characterize the interdot hopping. It should
be noted that the Rashba interaction brings about a spin dependent
phase factor $\sigma\varphi$.
\par
We now proceed to study the electronic transport through this QD
ring. By means of the Green function technique, at zero temperature
the electron current with a specific spin in an arbitrary lead, say
lead-1, can be expressed as\cite{Meir}
\begin{equation}
J_{1\sigma}=\frac{e}{h}\sum_{j'\sigma'}\int_{\mu_1}^{\mu_{j'}}
d\omega T_{1\sigma,j'\sigma'}(\omega),\label{current}
\end{equation}
where $T_{1\sigma,j'\sigma'}(\omega)=\Gamma_1
 G^r_{1\sigma,j'\sigma'}(\omega)\Gamma_{j'}G^a_{j'\sigma',1\sigma}(\omega)$
denotes the transmission probability between spin-$\sigma'$ electron
in lead-$j'$ and spin-$\sigma$ electron in lead-1. $\Gamma_j=2\pi
|V_{j\sigma}|^2\rho_j(\omega)$, associated with the density of
states in lead-j $\rho_j(\omega)$, can be usually regarded as a
constant if $\rho_j$ is a slow-varying function in the energy scale
as far as the electron transport is concerned.\cite{gamma} $G^r$ and
$G^a$, the retarded and advanced Green functions, are $6\times 6$
matrixes for the triple-QD ring. They have the relationship
$[G^r]=[G^a]^\dag$. From the equation-of-motion method, we obtain
the retarded Green function matrix,
\begin{eqnarray}
&&[G^r]^{-1}=\notag\\
&&\left[\begin{array}{cccccc} g_{1}^{-1} & -t_{1\uparrow}&-t_{3}e^{-i\phi}&0&r^*_1&0\\
  -t^*_{1\uparrow}& g_{2}^{-1}& -t_{2\uparrow}&-r^*_1&0&r^*_2\\
  -t_{3}e^{i\phi}&-t^*_{2\uparrow}&g_{3}^{-1}&0&-r^*_2&0 \\
  0&-r_1&0&g_{1}^{-1} & -t_{1\downarrow}&-t_{3}e^{-i\phi}\\
  r_1&0&-r_2&-t^*_{1\downarrow}& g_{2}^{-1}& -t_{2\downarrow}\\
  0&r_2&0&-t_{3}e^{i\phi}&-t^*_{2\downarrow}&g_{3}^{-1}
\end{array}\right]\notag
\end{eqnarray}
In the above expression, $g_j$ is the Green function of QD-j
unperturbed by the other QDs and in the absence of Rashba effect.
$g_{j}=[(z-\varepsilon_{j})+\frac{i}{2}\Gamma_j]^{-1}$ with
$z=\omega+i0^+$.
\par
As for the chemical potentials in the three leads, we fix $\mu_1=0$.
It is the reference point of energy of the system. And we let
$\mu_2=-\mu_3=eV/2$ with $V$ being a small bias voltage. Then the
net charge $J_{1c}$ and spin currents $J_{1s}$ in lead-1 are
respectively defined as $J_{1c}=J_{1\uparrow}+J_{1\downarrow}$ and
$J_{1s}=J_{1\uparrow}-J_{1\downarrow}$.
\par
Now we are ready to carry out the numerical calculation about the
spectra of the charge and spin currents in lead-1. To do this, we
choose the Rashba coefficient $\tilde{\alpha}=0.5$ which is
available under the current experimental circumstance\cite{Sarra}.
And the bias voltage is $eV=2t_0$ with $t_0$ being an appropriate
unit of energy.  In Fig.\ref{structure}(b) $J_{1c}$ and $J_{1s}$
versus the magnetic phase factor $\phi$ are shown. Besides, their
traces as a function of the QD level are shown in
Fig.\ref{structure}(c) and (d). The following interesting features
in these spectra are noteworthy. With the variation of the applied
magnetic field, the charge and spin currents oscillate out of phase.
In the vicinity of $\phi=(n+\frac{1}{2})\pi$, namely, the magnetic
phase factor is the odd multiple of $\pi/2$, $J_{1c}$ reaches its
maximum. Simultaneously, the spin current $J_{1s}$ just be very
close to a zero point. On the contrary, when $\phi=n\pi$ the
situation is just inverted, the maximum of $J_{1s}$ encounters the
zero of $J_{1c}$. This indicates that a striking pure spin current
can be implemented without an accompanying charge current. In
particular, such a pure spin current emerges even at the vicinity of
$\phi=0$, which implies that an applied magnetic field is not an
indispensable condition for the occurrence of the pure spin current.
\par
In Fig.\ref{structure}(c), the currents versus the QD levels are
shown in the absence of magnetic field. Apart from the pure spin
current at a specific value of $\varepsilon_0$, one can find the
more interesting phenomenon that the polarization direction of the
spin current can be inverted by the reversal of the bias. In
addition, as shown in Fig.\ref{structure}(d), when the coupling of
the QD ring to lead-3 is cut off, the spin current disappears,
though the Rashba interaction still exists. This means that the
three-terminal configuration is a necessary condition for the
occurrence of pure spin current.
\par
\begin{figure}
\centering \scalebox{0.4}{\includegraphics{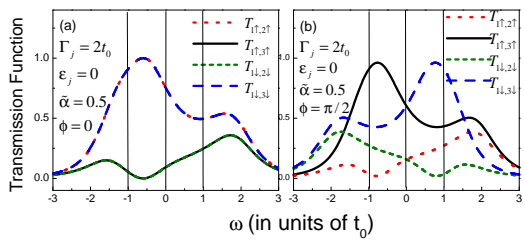}}\caption{
(Color online) The spectra of transmission functions
$T_{1\sigma,j\sigma}$($j=2,3$). In (a) no magnetic field is taken
into account; In (b) magnetic field is considered with
$\phi=0.5\pi$. \label{Transmission}}
\end{figure}

The calculated transmission functions are plotted in
Fig.\ref{Transmission}. They are just the integrands for the
calculation of the charge and spin currents, see Eq.(\ref{current}).
By comparing the results shown in Fig.\ref{Transmission}(a) and (b)
we can see that these transmission functions depend nontrivially on
the magnetic phase factor. At $\phi=0$, i.e., the zero magnetic
field case, the traces of transmission functions
$T_{1\sigma,2\sigma}(\omega)$ and
$T_{1\bar{\sigma},3\bar{\sigma}}(\omega)$ coincide with each other
very well. However, as shown in Fig.\ref{Transmission}(b), at
$\phi=\pi/2$ the four transmission function show distinct traces. In
this case a noticeable feature is the transmission between lead-1
and lead-2 is relatively suppressed, in comparison with
$T_{1\sigma,3\sigma}(\omega)$. Substituting such integrands into the
current formulae, and noting the opposite bias voltages of lead-2
and lead-3 with respect to lead-1, one can certainly arrive at the
results of the distinct charge and spin currents at $\phi=0$ and
$\pi/2$ respectively, as shown in Fig.\ref{structure}.

\par
The underlying physics being responsible for the spin dependence of
the transmission functions is the quantum interference, which
becomes manifest if we analyze the electron transmission process in
language of the Feynman path. First of all, we notice that the spin
flip arising from the Rashba interaction does not play a leading
role in causing the tunable spin and charge currents. To illustrate
this, we plot the spectra of the charge and spin currents in the
absence of the spin flip term(i.e., $r_l=0$) in
Fig.\ref{structure}(b). We can see that the corresponding results
coincide with the complete spectra very well. Therefore, to keep the
argument simple, we drop the spin flip term for the analysis of
quantum interference.
\par

We write $T_{1\sigma,2\sigma}=|\tau_{1\sigma,2\sigma}|^2$ by
introducing the transmission probability amplitude which is defined
as $\tau_{1\sigma,2\sigma}=\widetilde{V}^*_{1\sigma}
G^r_{1\sigma,2\sigma}\widetilde{V}_{2\sigma}$ with
$\widetilde{V}_{j\sigma}=V_{j\sigma}\sqrt{2\pi\rho_j(\omega)}$. The
transmission probability amplitude $\tau_{1\sigma,2\sigma}$ can be
divided into two terms, i.e.,
$\tau_{1\sigma,2\sigma}=\tau^{(1)}_{1\sigma,2\sigma}+\tau^{(2)}_{1\sigma,2\sigma}$,
where
$\tau^{(1)}_{1\sigma,2\sigma}=\frac{1}{D}\widetilde{V}^*_{1\sigma}
g_{1}t_{1\sigma}g_{2}\widetilde{V}_{2\sigma}$ and $
\tau^{(2)}_{1\sigma,2\sigma}=\frac{1}{D}\widetilde{V}^*_{1\sigma}
g_{1}t_{3}e^{-i\phi}g_{3}t^*_{2\sigma} g_{2}\widetilde{V}_{2\sigma}$
with $D=\det \{[G^r]^{-1}\}$. By observing the structures of
$\tau^{(1)}_{1\sigma,2\sigma}$ and $\tau^{(2)}_{1\sigma,2\sigma} $,
we can readily recognize that they just represent the two paths from
lead-2 to lead-1 via the two arms of the QD ring. The phase
difference between them is $\Delta\phi_{2\sigma}=[\phi-2\sigma
\varphi-\theta_3]$ with $\theta_j$ being the argument of $g_{j}$.
$T_{1\sigma,3\sigma}$ can be analyzed in a similar way. That is
$T_{1\sigma,3\sigma}=|\tau^{(1)}_{1\sigma,3\sigma}+\tau^{(2)}_{1\sigma,3\sigma}|^2$,
with
$\tau^{(1)}_{1\sigma,3\sigma}=\frac{1}{D}\widetilde{V}^*_{1\sigma}g_{1}
t_{3}e^{-i\phi}g_{3}\widetilde{V}_{3\sigma}$ and
$\tau^{(2)}_{1\sigma,3\sigma}=\frac{1}{D}\widetilde{V}^*_{1\sigma}g_{1}
t_{1\sigma}g_{2}t_{2\sigma}g_{3}\widetilde{V}_{3\sigma}$. The phase
difference between the two paths is
$\Delta\phi_{3\sigma}=[\phi-2\sigma \varphi+\theta_2]$. Using the
parameter values given in Fig.2, we can evaluate that
$\varphi\approx\pi/6$, and $\theta_2=\theta_3=-\frac{\pi}{2}$ at the
point of $\omega=0$. Thus, at $\phi=0$ we have
$\Delta\phi_{2\uparrow}=\pi/6$ and
$\Delta\phi_{2\downarrow}=5\pi/6$, which clearly proves that the
quantum interference between $\tau^{(1)}_{1\uparrow,2\uparrow}$ and
$\tau^{(2)}_{1\uparrow,2\uparrow}$ (electron with spin up) is
constructive. But $\tau^{(1)}_{1\downarrow,2\downarrow}$ and
$\tau^{(2)}_{1\downarrow,2\downarrow}$ of spin down electron are of
destructive interference. Moreover, we can get
$\Delta\phi_{3\uparrow}=5\pi/6$ and
$\Delta\phi_{3\downarrow}=\pi/6$. This indicates that the situation
of the quantum interference between
$\tau^{(1)}_{1\uparrow,3\uparrow}$ and
$\tau^{(2)}_{1\uparrow,3\uparrow}$ is just opposite to that between
$\tau^{(1)}_{1\downarrow,2\downarrow}$ and
$\tau^{(2)}_{1\downarrow,2\downarrow}$. In the case of
$\phi=\frac{1}{2}\pi$, by a simple evaluation we find that
$\Delta\phi_{2\uparrow/\downarrow}=(+/-)2\pi/3$ and
$\Delta\phi_{3\uparrow/\downarrow}=(-/+)\pi/3$. Accordingly,
$T_{1\sigma,j\sigma}$ does not depend on the spin index sensitively.
But the constructive interference leads to the nontrivial increase
of $T_{1\sigma,3\sigma}$, in comparison with $T_{1\sigma,2\sigma}$.
Up to now, the characteristics of the transmission functions as
shown in Fig.\ref{Transmission}, hence the tunability of charge and
spin currents, have been clearly explained by analyzing the quantum
interference between two kinds of paths via two different arms of
the QD ring. In the case of zero magnetic field, the fact that the
charge current is irrelevant to the reversal of the bias voltage can
also be understood, since the profile of $T_{1\sigma,2\sigma}$ is
the same as that of $T_{1\bar{\sigma},3\bar{\sigma}}$, as shown in
Fig.\ref{Transmission}(a).
\par
Now let us see what happens when lead-3 is removed from the QD-ring.
At $\Gamma_3=0$, $g_{3}$ blows up. This leads to
$|\tau_{1\sigma,2\sigma}^{(1)}|\ll |\tau_{1\sigma,2\sigma}^{(2)}|$,
which implies that QD-3 provides a sharply resonant path for
electron transmission. As a result, the conductance is mainly
determined by $\tau_{1\sigma,2\sigma}^{(2)}$. The other path
$\tau_{1\sigma,2\sigma}^{(1)}$ is only the trivial perturbation, and
no spin polarization comes up. In the absence of magnetic field,
$|\tau_{1\sigma,2\sigma}^{(1)}|$ is not relevant to the electron
spin. Therefore, in Fig.\ref{structure}(d) we obtain the vanishing
spin current.
%\section{summary}
\par
In summary, in the present triple-QD ring, the local Rashba
interaction provides a spin-dependent A-B phase difference. The
three-terminal configuration balances the electron transmission
probabilities via two different arms of the QD ring. The variation
of the magnetic field strength and the QD level can adjust the phase
difference between the two kinds of Feynman paths on an equal
footing. Thus the spin-dependence of the electron transmission
probability can be controlled by altering the exerted magnetic field
or the QD levels. Furthermore, with a specific bias it is possible
to obtain the tunable charge and spin currents. Before ending our
work, we should remark briefly on the effect of the electron
interaction which we have ignored. The electron interaction can
cause the correlated electron transport, such as the Kondo
effect.\cite{Kondo} However, our results are obtained away from the
Kondo regime. To incorporate the Hubbard term into the Hamiltonian,
and by using the second-order approximation to truncate the Green
function equation, we have calculated the spectra of the charge and
spin currents. We find that although the Hubbard $U$ splits QD
levels, hence the resonant peaks are divided into two groups, the
tunable charge and spin currents remain.

\pagebreak

\clearpage
%\section{\protect\bigskip\ {\protect\large FIGURES}}

\bigskip

\end{document}